\newcommand{\bzero}{\mathbf 0}

\newcommand{\br}{\mathbf r}
 
\newcommand{\bx}{\mathbf x} 
\newcommand{\bz}{\mathbf z}

 \newcommand{\bF}{\mathbf F}

\newcommand{\mO}{\mathcal O}

\newcommand{\dif}{{\rm d}}
\newcommand{\me}{{\rm e}}
\newcommand{\mi}{{\rm i}}

\newcommand{\rT}{{\rm T}}

\newcommand{\rc}{{\rm c}}

 \newcommand{\rR}{{\rm R}}
\newcommand{\rs}{{\rm s}}
\newcommand{\rt}{{\rm t}}

\newcommand{\brho}{\boldsymbol \rho}

\newcommand{\bOmega}{\boldsymbol \Omega}

\documentclass[preprint,12pt,authoryear]{elsarticle}

\usepackage{amssymb}
\usepackage{amsthm}

\usepackage{lineno}

\usepackage{url,hyperref,microtype,subcaption}
\usepackage[onehalfspacing]{setspace}

\usepackage{amsmath,amsfonts,amssymb}
\usepackage{graphics,graphicx,color}

\usepackage{nicefrac}

\usepackage{threeparttable}
\usepackage[table]{xcolor}
\usepackage{layout}

\usepackage{hyperref}
\hypersetup{
    colorlinks=true,
    linkcolor=blue,
    filecolor=magenta,      
    urlcolor=cyan,
    citecolor=blue}

\journal{Icarus}

\begin{document}


\begin{frontmatter}



\title{Imaging Venus' surface at night in the near-IR from above its clouds: 
       New analytical models for the effective spatial resolution, illustrated with new Parker Solar Probe data}


\author[inst1]{Anthony B. Davis} 

\affiliation[inst1]{
            organization={Jet Propulsion Laboratory, California Institute of Technology},
            addressline={4800 Oak Grove Drive}, 
            city={Pasadena},
            postcode={91109}, 
            state={CA},
            country={USA}
            }

%

\begin{abstract}
There are a handful of spectral windows in the near-IR through which we can see down to Venus’ surface on the night side of the planet. 
The surface of our sister planet has thus been imaged by sensors on Venus-orbiting platforms (Venus Express, Akatsuki) and during fly-by with missions to other planets (Galileo, Cassini). 
The most tantalizing finding, so far, is the hint of possible active volcanism. 
However, the thermal radiation emitted by Venus’ searing surface (c. 475$^\circ$C) has to get through the opaque clouds between 50 and 70 km altitude, as well as the sub-cloud atmosphere. 
In the clouds, the light is not absorbed but scattered, indeed, many times. 
This results in blurring the surface imagery to the point where the smallest discernible feature is roughly 100~km in size, full-width half-max (FWHM), and this has been reproduced using numerical models. 
We propose a new analytical modeling framework for predicting the width of the atmospheric point-spread function (APSF) that determines the effective resolution of surface imaging from space.
Our best estimates of the APSF width for the 1-to-1.2~$\mu$m spectral range are clustered around 130~km FWHM.
Interestingly, this is somewhat larger than the accepted value of $\approx$100~km based on both visual image inspection and detailed numerical simulations.
Lastly, we apply the new modeling framework to the fly-by imaging by the Parker Solar Probe in a somewhat shorter wavelength band.
\end{abstract}




\end{frontmatter}

\section{Introduction \& Overview}
\label{sec:intro}

Nighttime imaging by near-IR (NIR) sensors, either orbiting Venus (Venus Express, Akatsuki) or during fly-bys (Galileo, Cassini), have provided glimpses at the planet's surface through a handful of spectral windows between CO$_2$ absorption bands, revealing the possible presence of active volcanism \citep{smrekar2010recent,mueller2017search,byrne2020estimates}. 
Surface-emitted thermal radiation must however get through optically thick light-scattering clouds between 50 and 70 km in altitude \citep[e.g.,][and references therein]{titov2018clouds}.
Therefore, the effective spatial resolution (ESR) of the surface imaging is poor, at $\approx$100 km. 
This is a long-accepted fact based on visual inspection of the observations. followed by an intuitive geometrical argument \citep{Moroz2002} or detailed numerical simulations \citep{Hashimoto2001Elucidating,basilevsky2012geologic}.

\begin{figure}[htbp]
\centering
\includegraphics[width=5.5in]{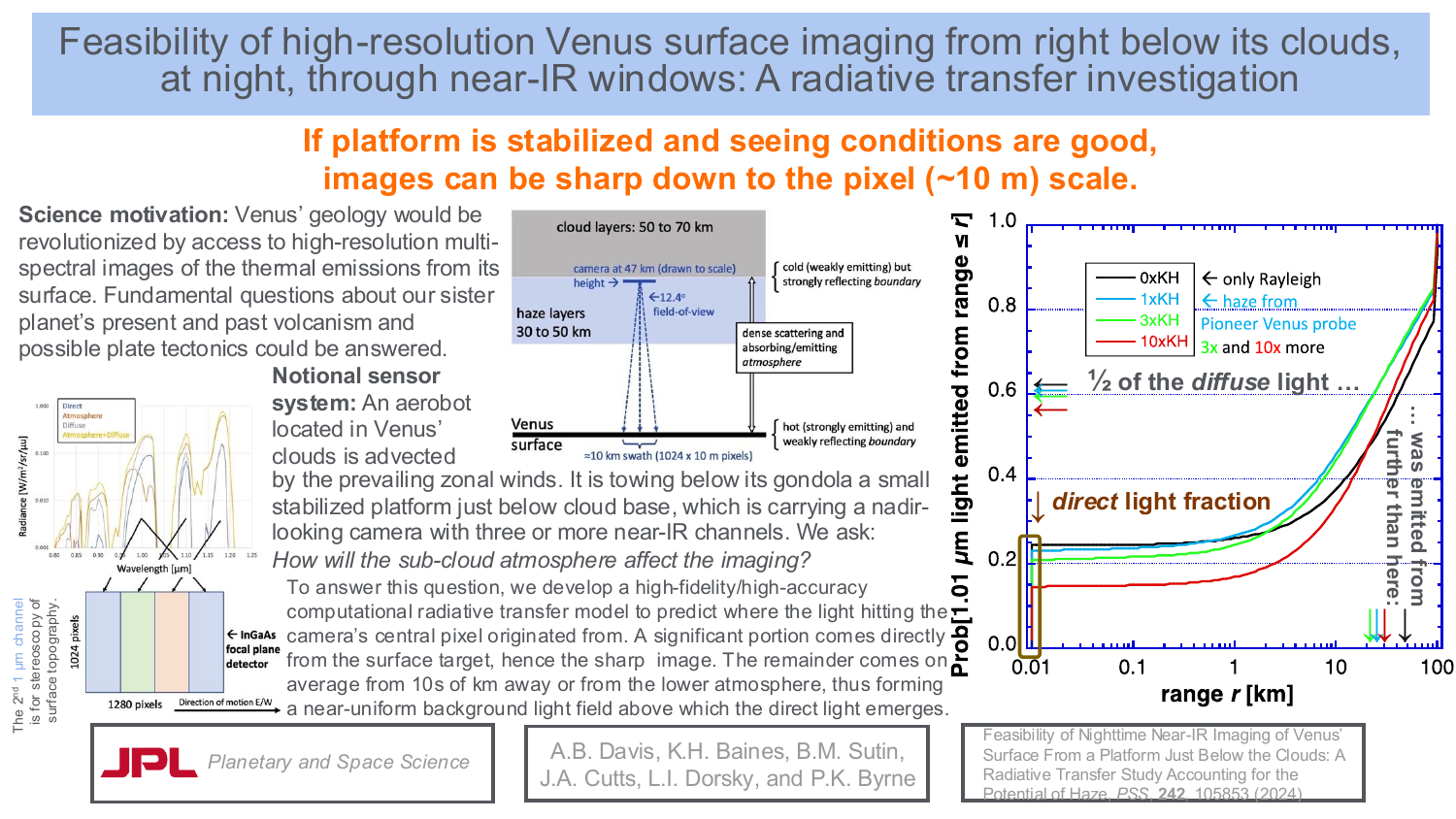}
\caption{Graphical summary of the investigation by \cite{DavisEtAl_PSS_2024} of the feasibility of sharp nighttime NIR imaging of Venus' thermally glowing surface from a platform located just below the clouds.}
\label{fig:graphical_abst_PSS}
\end{figure}

In a recent study \citep{DavisEtAl_PSS_2024}, we used numerical estimates of the atmospheric point-spread function (APSF) to show that sharp nighttime surface imaging is feasible \emph{from just below} Venus' clouds; see Fig.~\ref{fig:graphical_abst_PSS}.
Here, we present a family of analytical models for the APSF that predict it, or its at least its width, in closed form for space-based imaging.
However, the new model suggests somewhat larger values than currently accepted, specifically, we find values $\approx$130~km, full-width half max (FWHM) for the 1-to-1.2~$\mu$m NIR spectral subrange on which \cite{DavisEtAl_PSS_2024} focused their attention.

In the following Section, the clouds are modeled in a preliminary fashion as a thin diffusing plate at an altitude of $\approx$50~km, which is cloud-base height, and sub-cloud propagation is approximately ballistic, i.e., dominated by direct transmission.
In Section~\ref{sec:diffusion_theory}, we adopt a more realistic representation of clouds as a multiple scattering optical medium, and apply closed-form expressions from a diffusion theoretical model developed originally for Earth observation \citep{davis2002space}.
In Section~\ref{sec:PSP_flyby}, we extend the new modeling framework for application to flyby observations by the Parker Solar Probe (PSP) at more visible (VIS) than NIR wavelength.
Section~\ref{sec:conclusion} summarizes our findings.

\section{Cloud layer as a diffusing plate at 50~km altitude}
\label{sec:plate_clouds}

\begin{figure}[htbp]
\centering
\includegraphics[width=5.4in]{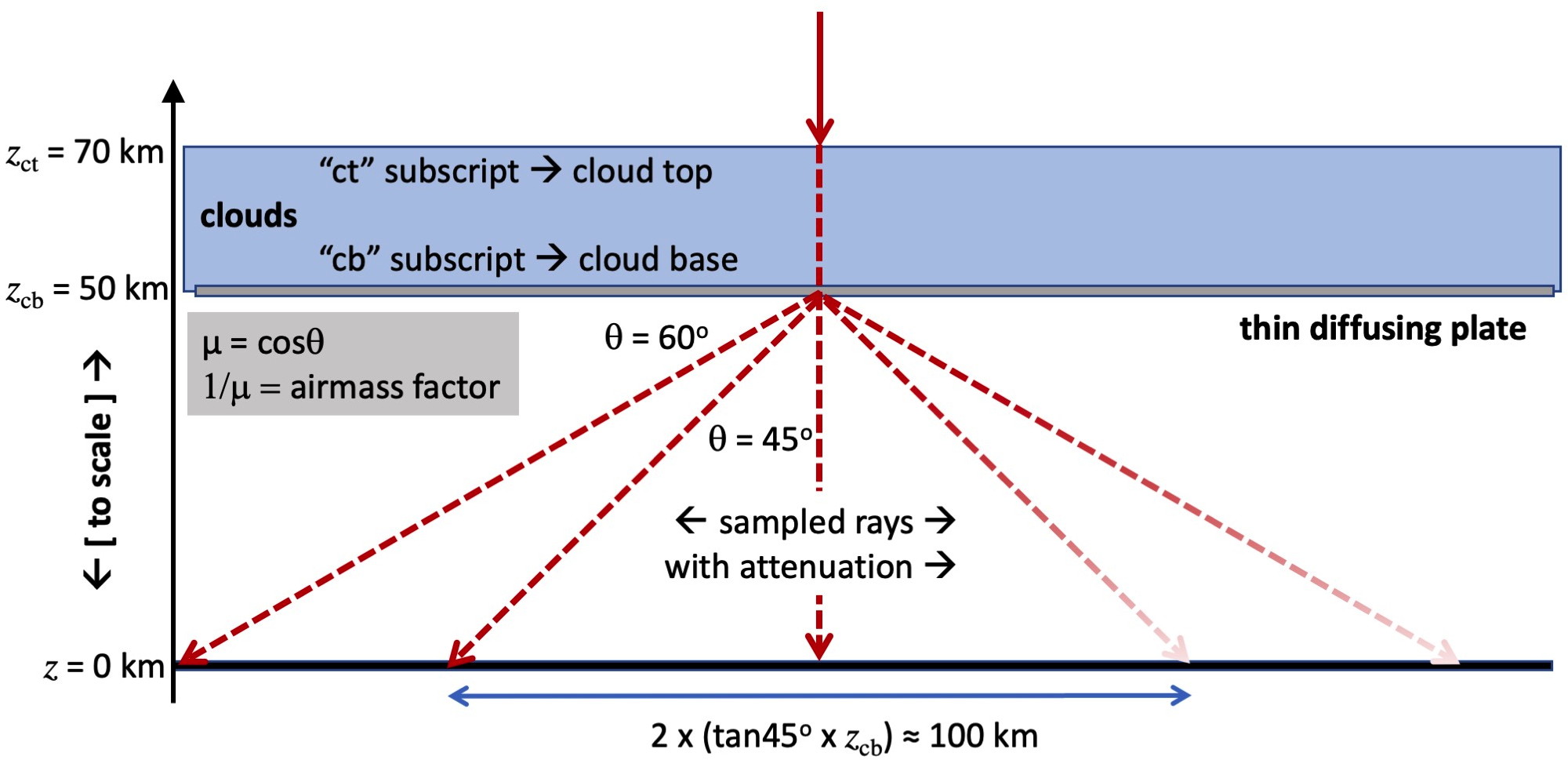} \\
\includegraphics[width=5.4in]{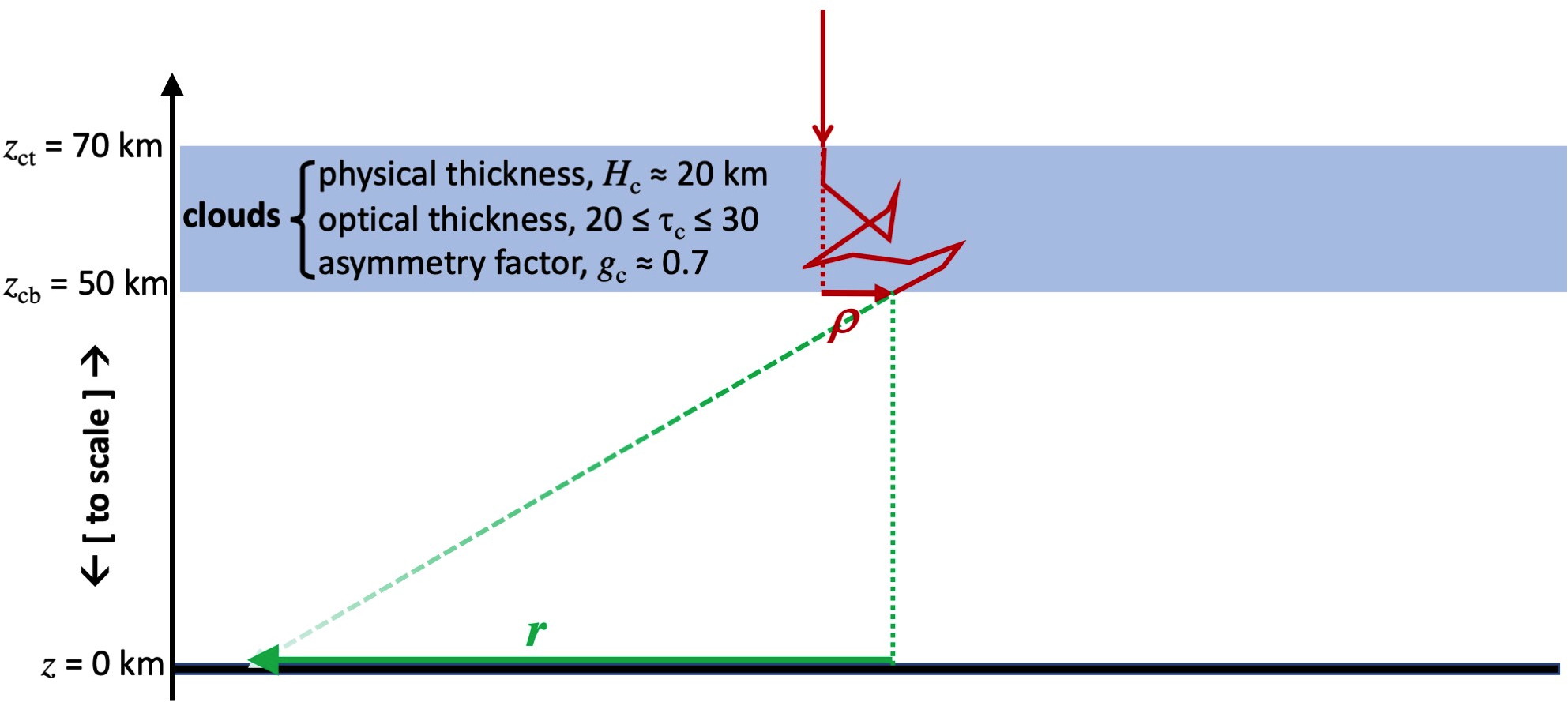}
\caption{Schematic for the estimation of the effective resolution of surface features achievable from space. 
{\bf Top panel:} As a first-order approximation, the clouds act as a diffusing plate; thus, the smallest resolvable feature depends only on cloud-base height (left-hand side), which is $\approx$50~km. 
Factoring in the extinction across the sub-cloud Rayleigh-scattering atmosphere (right-hand side), the effective resolution becomes somewhat sharper since near-nadir rays weigh in more heavily on average. 
{\bf Bottom panel:} Conceptualization of the cloud layer as a diffusing plate is upgraded to a strongly scattering medium of finite thickness where transmitted light is dispersed away from the horizontal position of the source by a predictable RMS distance. Effective surface resolution from above the clouds is then the RMS sum of two horizontal transport mechanisms: multiple scattering inside the cloud, and attenuated ballistic propagation below the cloud. 
See text for further details.}
\label{fig:AppB_Figs}
\end{figure}

Figure~\ref{fig:AppB_Figs} (top panel) is a schematic of the simplest possible conceptual model where Venus' clouds act like a thin diffusing plate 
Through such a plate, one can clearly see the shape of things on the opposite side only if they are up next to it.
On Venus, the objects of interest are 50~km below, so they cannot be distinguished clearly.
In fact, the minimum size of any distinguishable feature is determined by a directionally averaged incoming angle, thus leading to the $\sim$100~km minimum size of a distinguishable surface feature.

\begin{table}[htbp]
\caption{{\it From left to right:} 
Venus' sub-cloud atmosphere's optical thickness for Rayleigh scattering at three NIR wavelengths of interest, being CO$_2$ absorption windows. 
Total optical thickness, including an empirical band-wide estimate of gaseous absorption impacts. Predictions for the RMS diameter $2r_\text{RMS}$ (and FWHM) of the APSF for our three models for the impact of clouds: 
(1) clouds as a thin diffusing plate; 
(2) asymptotic diffusion theory; 
(3) with pre-asymptotic corrections. In all cases, the sub-cloud propagation is modeled as ballistic, with the appropriate attenuation; see Eq.~(\ref{eqn:RMS_tantheta}).
For Models 2 and 3, a Gaussian shape is assumed for the APSF, hence FWHM = 1.18$\, \times \, 2r_\text{RMS}$.
For Model 1, the APSF's shape is not Gaussian, and the expression in (\ref{eqn:FWHM_ballistic}) is used for the FWHM.}
\begin{center}
\begin{tabular}{|l|cc|ccc|}
\hline
$\lambda$ [$\mu$m] & $\tau_\rR$ & $\tau_\rt$ & Model 1 [km] & Model 2 [km] & Model 3 [km] \\
\hline
\hline
1.01         & 1.43  & 1.43  & 107 (114)  & 111 (131)  & 113 (133) \\
\hline
1.10         & 1.09  & 1.64  & 100 (105)  & 106 (125)  & 108 (127) \\
\hline
1.18         & 0.77  & 1.54  & 103 (108)  & 108 (127)  & 110 (130) \\
\hline
\end{tabular}
\end{center}
\label{tab:tau_values}
\end{table}

A closer look at the right-hand side of the top panel in Fig.~\ref{fig:AppB_Figs} reminds us that the sub-cloud space is not transparent. Directly-transmitted incoming light is extinguished across the medium by an amount that depends only on its optical thickness and the incoming zenith angle $\theta$.
Ruling out the possibility of haze, for simplicity, optical thickness is the Rayleigh scattering value $\tau_\rR$ listed in Table~\ref{tab:tau_values}.
Probability of direct transmission is then $\exp(-\tau_\rR/\mu)$ where we recognize the airmass factor $1/\mu$, with $\mu = \cos\theta$.

Now we can estimate the mean and root-mean square (RMS) value of the horizontal transport distance across the medium, namely, $r = z_\text{cb}\tan\theta$ where $z_\text{cb}$ is identified as the cloud base height in Fig.~\ref{fig:AppB_Figs}, which are drawn to scale.
Specifically, we have
\begin{equation}
\label{eqn:mean+RMS_tantheta}
\langle r^q \rangle = z_\text{cb}^q \, \langle \tan^q\theta \rangle 
                    = z_\text{cb}^q \int\limits_0^1 \left( \frac{1}{\mu^2}-1\right)^{q/2} \me^{-\tau_\rR/\mu} \dif\mu
                                    \left/ \left[ \me^{-\tau_\rR}-\tau_\rR\Gamma(0,\tau_\rR) \right] \right. \!,
\end{equation}
where $q = 1,2$ yield respectively the 1$^\text{st}$ and 2$^\text{nd}$ moments of $r$, the modulus of the random horizontal vector $\br = (r\cos\phi,r\sin\phi)$ with $\phi$ being the azimuthal angle.
In short, (\ref{eqn:mean+RMS_tantheta}) yields the mean $\langle r \rangle$ of $r$ and the variance $\langle \br^2 \rangle$ of $\br$ since $\langle \br \rangle$ vanishes by rotational symmetry.
From the latter quantity, we can obtain $r_\text{RMS} = \sqrt{\langle \br^2 \rangle}$.
The normalization factor in the denominator of (\ref{eqn:mean+RMS_tantheta}) is the result of the integral in the numerator when $q = 0$; therein, $\Gamma(a,x)$ is the incomplete Euler Gamma function.

For $q = 1$, the integration is performed numerically, but for $q = 2$, we have a closed-form expression:
\begin{equation}
\label{eqn:RMS_tantheta}
\langle \br^2 \rangle = z_\text{cb}^2 \, \langle \tan^2\theta \rangle 
                      = z_\text{cb}^2 \left[ \me^{-\tau_\rR}\left( 1/\tau_\rR-1 \right)+\tau_\rR\Gamma(0,\tau_\rR) \right]
                                      \left/ \left[ \me^{-\tau_\rR}-\tau_\rR\Gamma(0,\tau_\rR) \right] \right. \!.
\end{equation}
For the three finite values of $\tau_\rR$ in Table~\ref{tab:tau_values}, we find $2 r_\text{RMS} = 2\sqrt{\langle \br^2 \rangle}$ = 107, 120 and 140~[km], respectively for $\lambda$ = 1.01, 1.10 and 1.18~[$\mu$m] when $z_\text{cb}$ = 50~km.
For $q$ = 1, numerical quadrature of the integral in (\ref{eqn:mean+RMS_tantheta}) yields respectively $2 \langle r \rangle$ = 88, 98 and 112~[km].

Another popular measure of spatial resolution using the APSF is the full-width half-max (FWHM) distance.
Denoting the extinction coefficient as $\sigma_\rR = \tau_\rR/z_\text{cb}$, we identify the azimuthally-invariant APSF($r$) in (\ref{eqn:mean+RMS_tantheta}), as $\propto \exp\left(-\sigma_\rR\sqrt{z_\text{cb}^2+r^2}\right)$, that is, $\exp\left(-\tau_\rR/\mu_r\right)$.
In this case, 
\begin{equation}
\label{eqn:FWHM_ballistic}
\text{FWHM} = 2 z_\text{cb} \sqrt{(2\ln2)\tau_\rR+(\ln2)^2}/\tau_\rR, 
\end{equation}
and the three values of $\tau_\rR$ in Table~\ref{tab:tau_values} yield 114, 134 and 167~[km].

We note that, in the limit $\tau_\rR \to 0$ that is implicit in the left-hand side of Fig.~\ref{fig:AppB_Figs} (upper panel), both mean and variance in (\ref{eqn:mean+RMS_tantheta}) diverge due to the grazing view angles, $\theta \approx \pi/2$. 
The FWHM parameter in (\ref{eqn:FWHM_ballistic}) also becomes infinite for the same reason.
Therefore, in the absence of sub-cloud extinction, we need to estimate the effective surface resolution by choosing a typical value of $\theta$ and injecting it into $r = z_\text{cb}\tan\theta$.
We illustrate the natural choice of 60$^\circ = \cos^{-1}\langle \mu \rangle$, yielding 173~km, which seems too big.
Following \cite{Moroz2002}, we can rationalize the choice of 45$^\circ$, leading to the more reasonable estimate of $2 z_\text{cb}$ = 100~km resolution side-to-side.

So far, we have only accounted for extinction from Rayleigh scattering, which is a reasonable approximation for the shortest NIR wavelength under consideration (i.e., $\lambda \approx$ 1~$\mu$m).
For the longest ($\lambda \approx$ 1.2~$\mu$m), we cannot neglect absorption by gases (both CO$_2$ and H$_2$O contribute).
\cite{DavisEtAl_PSS_2024} did that (in their Appendix~A) using an \emph{effective} SSA$_\text{g} < 1$ in the atmospheric column, which was empirically estimated to be $\approx$1/2 for $\lambda$ = 1.18~$\mu$m and, for $\lambda$ = 1.10~$\mu$m, it was $\approx$2/3.
Now, using \emph{total} optical thickness $\tau_\rt = \tau_\rR/$SSA$_\text{g}$ in (\ref{eqn:RMS_tantheta}), we update 2$r_\text{RMS}$ in (\ref{eqn:RMS_tantheta}) to 100 and 103~km, respectively for $\lambda$ = 1.10 and 1.18~$\mu$m, while the FWHM in (\ref{eqn:FWHM_ballistic}) are reduced to 105 and 108~km.

These analytical estimates of the FHWM of the atmospheric spatial filter based on APSF moments are slightly larger than the $\sim$100~km value estimated numerically by \cite{Hashimoto2001Elucidating} and \cite{basilevsky2012geologic}, who account for multiple scattering in the cloud layer.
They follow, however, from a simple APSF model based only on attenuated ballistic propagation of the NIR light emanating from the surface across the sub-cloud optical medium.

\section{Cloud layer as a multiple scattering optical medium}
\label{sec:diffusion_theory}

We now turn to the lower panel of Fig.~\ref{fig:AppB_Figs} where we again upgrade our APSF model by relaxing the assumption of a diffusing plate with virtually no thickness to mimic the blurring effect of the cloud layer.
Instead, we now have two layers with finite optical and physical thicknesses that each disperse according to their individual APSFs the ``reciprocal'' (time-reversed) light that starts at a single point source at the upper boundary of the top layer.
Conversely, light emanating from a whole region at the lower boundary of the bottom layer propagates to the interface between the two layers, and a single pixel from the space-based sensor receives light from a whole region of that interface.
What is the size of the primary source region at the surface?
Conversely, what is the size of the spot of time-reversed light at the surface?
Either way, that will be the effective spatial resolution of the space-based sensor for features on the planetary surface.

In this problem, the most useful measure of APSF width is the RMS radius that we continue to denote $r_\text{RMS} = \sqrt{\langle \br^2 \rangle}$ for the sub-cloud layer, and introduce $\rho_\text{RMS} = \sqrt{\langle \brho^2 \rangle}$ for the cloud layer.
These RMS radii will both scale as the physical thickness of the layer, 50~km for the sub-cloud layer (denoted $z_\text{cb}$ in Fig.~\ref{fig:AppB_Figs}) and $\approx$20~km for the cloudy layer (denoted $H_\rc$ in the lower panel of Fig.~\ref{fig:AppB_Figs}).
Our best estimates for $r_\text{RMS}$ using the attenuated ballistic propagation model that we have used so far for the sub-cloud APSF are: $r_\text{RMS}$ = 107, 100 and 103~[km] for $\lambda$ = 1.01, 1.10 and 1.18~[$\mu$m], respectively.

We are now dealing with the sum of two \emph{independent} random horizontal vectors $\br$ and $\brho$, both with vanishing means.
We can therefore invoke the additivity property of the so-called ``cumulant'' moments \citep{renyi1970probability}.
In this case, only the first two are of interest.
The means yield $\langle \brho+\br \rangle = \langle \brho \rangle + \langle \br \rangle = \bzero$.
The variances yield
\begin{equation}
\langle (\brho + \br)^2 \rangle = \langle \brho^2 \rangle + \langle \br^2 \rangle + 2\langle \brho\cdot\br \rangle = \langle \brho^2 \rangle + \langle \br^2 \rangle,
\label{eqn:Var_sum}
\end{equation}
where $\langle \br^2 \rangle$ was estimated in Section~\ref{sec:plate_clouds}.
For the new quantity, $\rho_\text{RMS}$, we can use outcomes from the theoretical investigation by \cite{davis2002space} of the Green functions of optically thick clouds for transmitted light at non-absorbing wavelengths.
In sharp contrast with our present ballistic propagation model for the sub-cloud medium, the authors used the diffusion approximation where light propagates across the medium in the manner of a random walking particle, specifically, many small steps in random directions.
In terrestrial as well as Venusian clouds, the particulates are commensurate or larger than the NIR wavelengths of interest here \citep{KH_1980m,ragent1985particulate}
They are therefore have quite strongly forward-scattering phase functions, with asymmetry factors $g_\rc \approx 0.7$.
Specifically, \cite{davis2002space} show that $\rho_\text{RMS} \approx \sqrt{2/3}\,H_\rc$, irrespective of the scattering phase function in the limit of large ``scaled'' optical thickness, i.e., $\tau_\rt = (1-g_\rc)\tau_\rc \gg 1$.
They also computed the correction factor for finite values of $\tau_\rt$, namely, 6 to 9, based on the numbers in Fig.~\ref{fig:AppB_Figs}'s lower panel.
That correction factor is $\sqrt{1+(3+4X)/2X(1+X)}$ where $X = \tau_\rt/2\chi$ (cf. \S\ref{sec:diffuse_src}), where $\chi$ can be set to 2/3 \cite[among others]{meador1980two}.

Combining our predictions for the sub-cloud layer and those of \cite{davis2002space} for the cloudy layer according to (\ref{eqn:Var_sum}), we find that the overall APSF has an 2$\times$RMS size of $2\sqrt{\langle\brho^2\rangle+\langle\br^2\rangle}$ = 111, 106 and 108~[km], respectively, for $\lambda$ = 1.01, 1.10 and 1.18~[$\mu$m]. 
Neglecting in-cloud absorption by gases is justified at the longer wavelengths (where it matters in the sub-cloud layer) because molecular densities have decreased to a very small fraction of their near-surface values. 
Davis and Marshak's small correction for the range of values assumed for $\tau_\rc$ in the lower panel of Fig.~\ref{fig:AppB_Figs} is an increase of $\approx$16\% in $\rho_\text{RMS}$, assuming the mid-range value of $\tau_\rc \approx 25$, hence $(1-g_\rc)\tau_\rc$ = 7.5.
This modest boost in $\rho_\text{RMS}$ leads to 113, 108 and 110~[km], respectively, for (2 RMS) = $2\sqrt{\langle\brho^2\rangle+\langle\br^2\rangle}$, which is always dominated by the latter sub-cloud contribution.

To convert our estimates of the RMS radius into the more widely used FWHM in spatial filtering, we assume a Gaussian shape for the APSF.
This leads to FWHM = $\sqrt{2\ln\!2}\times$(2 RMS) $\approx 1.1774\times$(2 RMS).
The FWHM values associated with our above best estimates of (2 RMS) are 133, 127, and 130~[km], respectively, for $\lambda$ = 1.01, 1.10 and 1.18~[$\mu$m].
These values are significantly ($\approx$30\%) larger than previous estimates of $\approx$100~km by \cite{Hashimoto2001Elucidating} and \cite{basilevsky2012geologic} using forward Monte Carlo simulation schemes.
The discrepancy between ours and their predictions is then exacerbated since they conclude their analyses by stating (without quantitative evidence) that features as small as $\approx$50~km may be detectable.

\section{Extension to flyby imaging by the Parker Solar Probe}
\label{sec:PSP_flyby}

\begin{figure}[htbp]
\begin{center}
\includegraphics[width=0.96\linewidth]{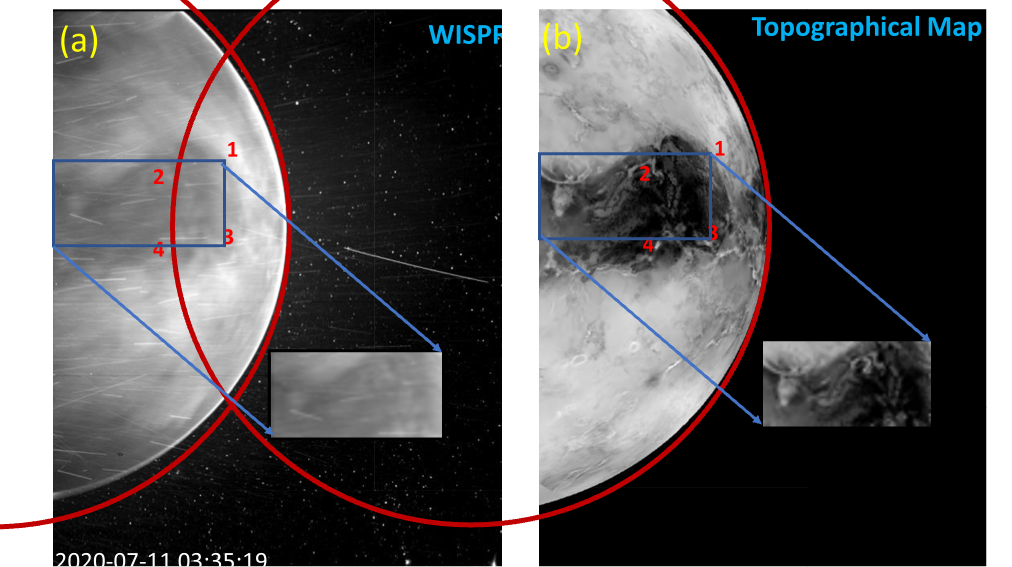}
\caption{{\bf (a) WISPR/PSP image of the nighttime side of Venus.}
{\bf (b)} Venus topography map from Magellan projected on to the side-viewed sphere.}
\label{fig:Venus_WISPR-PSP}
\end{center}
\end{figure}

\subsection{Empirical estimate of the effective surface footprint of WISPR cameras}

The Wide-Field Imager for Solar PRobe (WISPR) \cite{vourlidas2016wide} cameras on the PSP \cite{wood2022parker} are sensitive to light that is heavily weighted at around 0.8~$\mu$m in wavelength.
PSP has had seven Venus gravity assists, each one an opportunity to turn WISPR on to the nightside of Venus.
Figure~\ref{fig:Venus_WISPR-PSP}a shows a raw WISPR image captured during the 3rd gravity assist on July 11, 2020.
Figure~\ref{fig:Venus_WISPR-PSP}b displays version 2 of the Magellan Global Topographic Data Record \citep{FordPettengill1992} where the grayscale is inverted to facilitate comparison with the WISPR image, hence bright areas mean low altitude and dark ones high altitude.
A rectangular sub-image is isolated in both cases as far as possible from the limb.
The effective WISPR pixel scale is estimated to be $\approx$27~km, based on the known radius of the planet (6052~km), which is commensurate with the 10~km resolution of the radar-based topography map.

\begin{figure}[htbp]
\begin{center}
\includegraphics[width=1\linewidth]{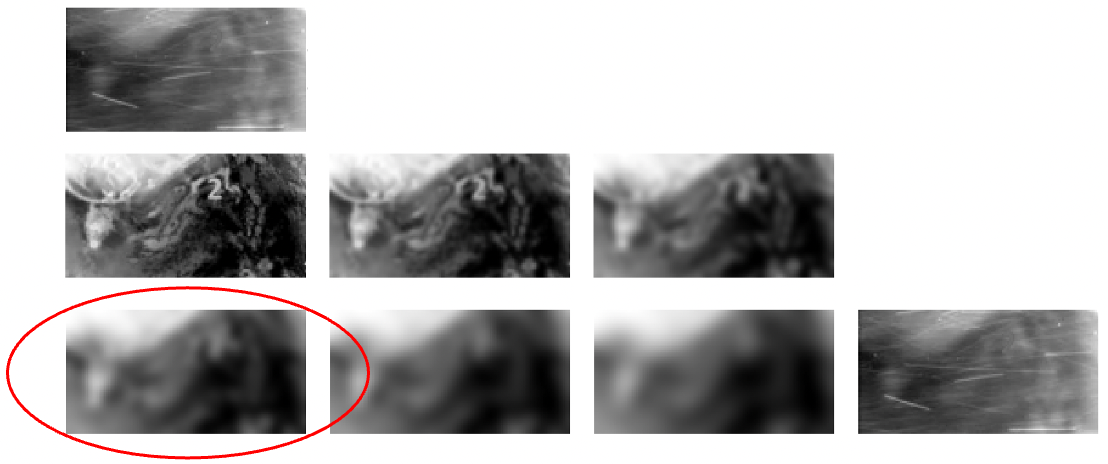}
\caption{Empirical estimate of the effective spatial resolution of WISPR cameras when imaging Venus' surface at 0.8~$\mu$m, as determined by the APSF and described in main text.}
\label{fig:Venus_WISPR_footprint}
\end{center}
\end{figure}

To produce Fig.~\ref{fig:Venus_WISPR_footprint}, the WISPR subimage's dynamic range was stretched to match that of the (inverted) topography.
It is then displayed in the top and rightmost images in the montage of 8 small panels.
The sequence of 6 small images in the remaining 2-by-3 array are produced by smoothing the sharp topography sub-image from Fig.~\ref{fig:Venus_WISPR-PSP} using gaussian kernels of widths 0, 2, 4, 6, 8, and 10 pixels, where 0 means no smoothing.
Visual comparison of the smoothed topography images and the (stretched) WISPR image, ignoring the artifacts (white streaks), shows that the 6-pixel smoothing kernel is the best match.
Therefore our empirical estimate of the effective surface footprint of WISPR is 6$\times$27 = 162~km.
As expected from the increased scattering to be expected in the sub-cloud Rayleigh layer, this is larger than all estimates for the longer NIR wavelengths.

\subsection{Theoretical estimate of WISPR's anticipated spatial resolution}
\label{sec:WISPR_footprint_theory}

WISPR cameras filter light through a band concentrated at 0.8~$\mu$m. 
That does not change much from the previously considered NIR wavelengths as to the optical properties of the clouds since the particles of sulfuric acid are relatively large in the densest layers compared to the wavelengths of interest so far, and even more so at 0.8~$\mu$m.
This partially explains why, like on Earth, Venus' clouds are basically white: the particles scattering according to geometric optics across the visible to NIR spectrum.
However, a major difference arises in the sub-cloud molecular (Rayleigh scattering) atmosphere.
Going from $\tau_\rR$ = 1.43 at 1.01~$\mu$m (cf. Table~\ref{tab:tau_values}) to 1.43$\times(0.80/1.01)^{-4} \approx 3.6$ at 0.8~$\mu$m, makes a major difference in the radiative transfer regime.
Rather than at least partially ballistic propagation, hence the model radial dispersion in Eq.~(\ref{eqn:RMS_tantheta}), we need a multiple scattering model akin to \cite{davis2002space}, recalling that Rayleigh scattering is effectively isotropic ($g_\rR = 0$).
However, we now have a highly reflective (upper) boundary where the diffuse point source (of adjoint/time-reversed light) comes from and we ask how far the light will wander way from the sub-source point on the opposite (lower) boundary.
For consistency, we set the cloud layer albedo at \citep[for instance]{meador1980two}
\begin{align}
\label{eqn:albedo}
\alpha_\rc &= 1-T(\tau_\rt)\text{ where} \\
T(\tau_\rt) &= \frac{2\chi}{\tau_\rt + 2\chi}, \nonumber
\end{align}
which is 0.15 for the mid-range $\tau_\rc$ = 25 in the lower panel of Fig.~\ref{fig:AppB_Figs}.
Hence, $\tau_\rt = (1-g_\rc)\tau_\rc = (1-0.7) \times 25 = 7.5$ , leading to $\alpha_\rc$ = 0.85.
The radial dispersion for that scenario is computed in the diffusion (a.k.a. P$_1$) limit in \ref{sec:appendix_A}, yielding: 
\begin{equation}
\langle r^2 \rangle/H_\rR^2 = \frac{2}{3}\,
\frac{(1-\alpha_\rc)\tau_\rR^2 + 6 \tau_\rR\chi + 6(1+\alpha_\rc)\chi^2}
     {\tau_\rR\left( (1-\alpha_\rc)\tau_\rR + 2\chi \right)},
\label{eqn:r2_PSP}
\end{equation}
where $H_\rR = z_\text{cb} = 50$~km is the physical thickness of the sub-cloud Rayleigh scattering layer.
The right-hand side of (\ref{eqn:r2_PSP}) evaluates to $\approx$2.1.

For the cloud layer, we can also improve on the \cite{davis2002space} estimate of $\langle r^2 \rangle/H_\rc^2$, which can be identified with (\ref{eqn:r2_PSP}) in the limit $\alpha_\rc \to 0$.
Indeed, for this APSF computation, we are actually in a situation where the the ``adjoint'' (time-reversed) light beam is well-collimated when, coming from the WISPR camera, it impinges on the Venus cloud layer, and we'll assume normal incidence for simplicity.
Moreover, that layer is now sitting on top of a reflective lower layer of Rayleigh scattering with an optical thickness $\tau_\rR$~=~3.6, hence an albedo of $\alpha_\rR \approx 0.73$ according to (\ref{eqn:albedo}) and recalling that $g_\rR$ = 0, hence $\tau_\rt = \tau_\rR$.
This problem is also solved in \ref{sec:appendix_A} in the diffusion/P$_1$ approximation, leading to a complicated expression for $\langle \rho^2 \rangle/H_\rc^2$ that involves $\tau_\rc$, $g_\rc$, and $\alpha_\rR$. 
For the adopted values of those optical properties, it yields $\langle \rho^2 \rangle/H_\rc^2 \approx 1.3$.

In summary, we predict an effective surface footprint of $2\times\sqrt{\langle r^2 \rangle + \langle \rho^2 \rangle} = 2\times\sqrt{ 1.3 \times H_\rc^2 + 2.1 \times H_\rR^2} = 2\times\sqrt{ 1.3 \times 20^2 + 2.1 \times 50^2}$~[km] $\approx$ 152~km.
That is up from the estimate of $2\times\sqrt{ (2/3) \times (20^2 + 50^2)}$~[km] $\approx$ 88~km based on the simplified asymptotic diffusion model by \cite{davis2002space} and compares well with our empirical estimate of 162~km.

In short, accounting for reflections at the cloud/Rayleigh interface, the well-collimated incoming beam, and the finite optical thicknesses makes a significant difference.
This is of course all for diffuse light.
At the $\exp[-(\tau_\rc+\tau_\rR)] = \me^{-(25+3.6)} = 4\me$--13 level, there is no direct light left.
Even focusing on the sub-cloud Rayleigh scattering, we have $\exp[-\tau_\rR] = \me^{-3.6} \approx 0.1$ for the most favorable (vertical) propagation direction, which is far less than at the longer NIR wavelengths were direct transmission is a significant contribution to the sub-cloud transmission.
In fact, that non-negligible direct transmission in the NIR is at the core of the demonstration by \cite{DavisEtAl_PSS_2024} that it is possible to achieve camera-pixel sharpness when imaging the surface from just below the clouds.

\section{Summary}
\label{sec:conclusion}

We have used the atmospheric point-spread function (APSF) concept to predict from first principles the effective resolution of near-IR imagery of Venus surface features captured by space-based sensors, either orbiting or during a flyby.
Physically, the APSF quantities the transport of surface-emitted thermal radiation in the horizontal plane (i.e., across the imaging sensors' pixels) between the actual source and the surface point at which a given pixel is pointing.
Our APSF model is analytical, combining the two extreme propagation modes in radiative transfer: ballistic (below clouds) and diffusive (in clouds).
The clouds are first represented by a thin diffusing plate that does not add any lateral transport of the NIR light.
Next, they are given a finite thickness and opacity, thus enabling horizontal radiation transport inside the clouds, as modeled by \cite{davis2002space}, which is combined with the (dominant) contribution of the sub-cloud angular spread.
Our final estimates across the 1-to-1.2~$\mu$m wavelength range are narrowly clustered around $\approx$130~km full-width half-max (FWHM).
They thus verify analytically the accepted value of $\approx$100~km FWHM and suggest a somewhat larger number.
This excludes the lower $\approx$50~km FWHM values cited by several authors.

Finally, we extend our analytical modeling framework to estimate theoretically the effective surface resolution achieved by the Parker Solar Probe's WISPR cameras.
The outcome or $\approx$152~km agrees well with our empirical estimate of $\approx$162~km, thus validating the new model that is based in that case on the classic diffusion/P$_1$ approximation in radiative transfer but with possible reflections at the interface between the upper (cloud) layer and the lower (Rayleigh scattering) layer.

\appendix
\section{Two P$_1$ approximation models for multiple scattering in the WISPR/PSP application in \S\ref{sec:PSP_flyby}}
\label{sec:appendix_A}

\setcounter{equation}{0} \renewcommand{\theequation}{A.\arabic{equation}}
\setcounter{figure}{0} \renewcommand{\thefigure}{A.\arabic{figure}}
\setcounter{subsection}{0} \renewcommand{\thesubsection}{A.\arabic{subsection}}

\subsection{Previous investigations of nightside Venus surface imaging from space in the NIR}

There are two complimentary aspects of addressing the question of how well can we image a target through an optical medium in the presence of scattering.
Problem~1 is about contrast: How bright (or dark) is the target compared to its environment (that produces the ``background'' radiance field).
Problem~2 is about how much does the propagation across the medium degrade (or ``blur'') the transmitted radiance in the spatial and/or angular domains.
Problem~1 is about detectability: Is there enough contrast?
Problem~2 is about spatial and/or resolution: How big will a remotely observed point source be?

These problems have been investigated by many authors interested in imaging Venus' surface features either from above the clouds or, looking into future missions, from below the cloud layers.
Ever since \cite{baines2000detection}, there has been a sharp focus on nightside observations in the Near-IR (NIR), peering through ``windows'' in the CO$_2$ absorption spectrum.

From below the clouds, \cite{Moroz2002} developed a sophisticated two-stream RT model to study the contrast issue, while \cite{ekonomov2015resolving} addressed numerically the spatial resolution problem in Fourier space, i.e., using the atmospheric modulation transfer function (AMTF) rather than the APSF.
\cite{knicely2020evaluation} also investigated achievable spatial resolution from below the clouds using a closed-form analytical model used for blurring in CGI \citep{Ashikhmin2004Blurring}.
\cite{DavisEtAl_PSS_2024} used a detailed numerical model to address both Problems~1 and 2 for an imaging sensor just below the clouds.

From above the clouds, \cite{hashimoto2003observing} adapted the \cite{Moroz2002} model to study the contrast problem, while \cite{Hashimoto2001Elucidating} and then \cite{basilevsky2012geologic} used numerical models to evaluate the APSF.
Herein, we propose a new analytical modeling framework that uses the AMTF to get to physical-space properties of the APSF such as its FWHM, which is commonly used to determine the achievable spatial resolution.
In the main text, we proposed a ballistic propagation model for the sub-cloud medium applicable to the NIR where the optical thickness are at most $\mO(1)$. 
In the following, we develop a multiple scattering model in the diffusion/P$_1$ approximation, which is closely related to the two-stream model, to accommodate the higher optical thickness ($\approx$3.6) of the Rayleigh scattering medium.
We also propose in \S\ref{sec:collimated_src} a diffusion/P$_1$ approximation for the multiple scattering in the cloud layer itself.

\subsection{Definition of the APSF in radiative transfer (RT)}
\label{sec:RT_for_APSF}

Let $I(\bx,\bOmega)$ be the radiance field in [W/m$^2$/sr] at position $\bx = (x,y,z)^\rT$ propagating into direction $\bOmega$, a vector on the unitary sphere, with polar coordinates $(\theta,\phi)$.
It obeys the 3D RT equation,
\begin{equation}
\left[ \bOmega\cdot\nabla + \sigma \right]\,I = 
\omega\sigma\int\limits_{4\pi}p(\bOmega\cdot\bOmega^\prime)\,I(\bx,\bOmega^\prime)\,\dif\bOmega^\prime
+ Q(\bx,\bOmega),
\label{eqn:RTE}
\end{equation}
where $\bOmega\cdot\nabla$ is the directional derivative along $\bOmega$, $\sigma$ is the extinction coefficient in [1/m], $\omega$ is the single-scattering albedo (SSA), $p(\bOmega\cdot\bOmega^\prime)$ is the scattering phase function (PF) assumed to depend only on the scattering angle $\theta_\rs = \cos^{-1}\bOmega\cdot\bOmega^\prime$, and $Q(\bx,\bOmega)$ is the source term expressed in [W/m$^3$/sr].
On the left-hand side, we have two sinks of radiant energy in a small volume of transport space around $(\bx,\bOmega)$ via advection and extinction.
On the right-hand side, we have two sources: via in-scattering and via local emission.

Moreover, we have boundary conditions (BCs) at $z = 0$ and $z = H$, respectively, the upper and lower planes the define the plane-parallel optical medium, that define the incoming radiance.
We will use two cases in the remainder.

For one, we are interested in a unitary isotropic point source the origin on the upper boundary, which is also partially reflective (albedo $\alpha$) in a Lambertian (isotropic) pattern, which reads as
\begin{align}
\label{eqn:pt_sc_and_refl}
I_\text{bc}(x,y,\bOmega) &= \frac{1}{\pi}\,
                            \left[ \delta(x)\,\delta(y) + \alpha\,F_z^{(-)}(x,y,0) \right],
                            \text{ at }z = 0,\text{ for }\Omega_z > 0 \\
\label{eqn:pt_sc_and_refl_opposite}
I_\text{bc}(x,y,\bOmega) &= 0, 
                            \text{ at }z = H,\text{ for }\Omega_z < 0 
\end{align}
where $\delta(\cdots)$ is the Dirac delta function and 
\begin{equation}
F_z^{(\pm)}(x,y,z) = \int\limits_{0<\pm\Omega_z<1}|\Omega_z|\,I(x,y,z,\bOmega)\,\dif\bOmega
\label{eqn:F_z_up_down}
\end{equation}
is the hemispherical vector flux at $\bx$.
So, $F_z^{(-)}(x,y,0)$ is the hemispherical vector flux at $(x,y,0)$ flowing into the negative $z$ direction.
Also, we would set $Q(\bx,\bOmega) \equiv 0$ for this RT problem.
Note that, in this scenario, $I(\bx,\bOmega)$ is the \emph{total} (direct$+$diffuse) radiance.

For two, we use a homogeneous upper BC, that is, (\ref{eqn:pt_sc_and_refl}) is replaced by $I_\text{bc}(x,y,\bOmega) = 0 \text{ at }z = 0,\text{ for }\Omega_z > 0$.
Here, we assume a partially reflective lower boundary (again Lambertian): 
\begin{align}
I_\text{bc}(x,y,\bOmega) &= \,\frac{\alpha}{\pi}\,
                           \left[ \delta(x)\,\delta(y)\exp(-\sigma H) + F_z^{(+)}(x,y,H) 
                           \right], \nonumber \\
                         &\text{ at }z = H,\text{ for }\Omega_z < 0,
\label{eqn:pt_sc_opposite_refl}
\end{align}
where $\alpha$ is the surface albedo.
In this case, we still need a source of \emph{diffuse} radiance, and we take
\begin{equation}
Q(\bx,\bOmega) = \delta(x)\,\delta(y)\,\exp(-\sigma\,z)\,\omega\,\sigma\,p(\Omega_z),
\label{eqn:Q_for_collimated_src}
\end{equation}
in (\ref{eqn:RTE}), which models a unitary collimated point-source at the origin, thus, on the upper boundary, propagating straight down, i.e., into the positive $z$-direction that we denote as $\hat{\bz}$.
The required cosine of the scattering angle in (\ref{eqn:Q_for_collimated_src}) is therefore $\hat{\bz}\cdot\bOmega = \Omega_z$.
Here, the diffusely reflective boundary at $z = H$ is opposite the collimated point-source at $z = 0$ and the first term in (\ref{eqn:pt_sc_opposite_refl}) is the directly transmitted flux.

\subsection{Definition of the AMTF in the diffusion/P$_1$ approximation to RT}
\label{sec:RT_for_AMTF}

In both of the above cases, we are assuming uniform optical properties in the medium and in the BCs.
However, these are still 3D RT problems because of the localized sources.
That said, we can convert them into 1D RT problems by taking horizontal Fourier transforms, denoted with $\tilde{\null}$ and letting $(k_x,k_y)$ be the conjugate variables of $(x,y)$.
In the process, we are switching are focus from the APSF to its Fourier transform, the AMTF.
The RTE for $\tilde{I}(z,\bOmega)$ becomes
\begin{align}
\left[ \Omega_z\,\frac{\dif\:}{\dif z} + \sigma \right]\,\tilde{I} = 
&-\mi(k_x\,\Omega_x+k_y\,\Omega_y)\tilde{I} + 
\omega\sigma\int\limits_{4\pi}p(\bOmega\cdot\bOmega^\prime)\,\tilde{I}(z,\bOmega^\prime)\,\dif\bOmega^\prime \nonumber \\
&+ \tilde{Q}(z,\bOmega),
\label{eqn:FT_RTE}
\end{align}
where $(k_x,k_y)$ are now parameters of the model, much like optical properties, rather than independent variables.

We now recast our RT problems in the radiative diffusion limit.
Thus, we assume
\begin{align}
\label{eqn:I_diff}
\tilde{I}(z,\bOmega) =& \left[ \tilde{J}(z) + 3\,\bOmega\cdot\tilde{\bF}(z) \right] / 4\pi, \\
\label{eqn:p_diff}
p(\bOmega\cdot\bOmega^\prime) =& \left[ 1 + 3\,g\,\bOmega\cdot\bOmega^\prime \right] / 4\pi,
\end{align}
where $g = \int_{4\pi} \bOmega\cdot\bOmega^\prime\,p(\bOmega\cdot\bOmega^\prime)\,\dif\bOmega^\prime$ is the PF's asymmetry factor, and 
\begin{align}
\label{eqn:J_def}
\tilde{J}(z) =& \int\limits_{4\pi} \tilde{I}(z,\bOmega)\,\dif\bOmega, \\
\label{eqn:Fz_def}
\tilde{\bF}(z) =& \int\limits_{4\pi} \bOmega\,\tilde{I}(z,\bOmega)\,\dif\bOmega,
\end{align}
are respectively the scalar vector fluxes.

Applying $\int_{4\pi}[\cdots]$ and $\int_{4\pi}\bOmega[\cdots]$ to (\ref{eqn:FT_RTE}), using assumptions (\ref{eqn:I_diff})--(\ref{eqn:p_diff}) along with definitions (\ref{eqn:J_def})--(\ref{eqn:Fz_def}), we arrive at two coupled ordinary differential equations (ODEs) for $\{\tilde{J},\tilde{F}_z\}(z)$.
Letting $(\dif/\dif z)$ be denoted with primes, we have
\begin{align}
\label{eqn:conservation}
\tilde{F}_z^\prime(z) =& -\left[ \frac{k^2}{3(1-\omega\,g)\sigma} + (1-\omega)\sigma \right]\,\tilde{J}(z) + \tilde{Q}_J(z), \\
\label{eqn:constitutive}
\tilde{J}^\prime(z)/3 =& -(1-\omega\,g)\sigma\,\tilde{F}_z(z) + \tilde{Q}_{F_z}(z),
\end{align}
where $k^2 = k_x^2+k_y^2$ and
\begin{align}
\label{eqn:Q_J}
\tilde{Q}_J(z) &= \omega\sigma\,\exp(-\sigma z), \\
\label{eqn:Q_Fz}
\tilde{Q}_{F_z}(z) &= g\,\tilde{Q}_J(z) ,
\end{align}
if (\ref{eqn:Q_for_collimated_src}) is used, otherwise $\tilde{Q}_J(z) = \tilde{Q}_{F_z}(z) \equiv 0$.
Equation (\ref{eqn:conservation}) expresses the local conservation of radiant energy, where we note that the first term on the right-hand side correctly identifies horizontal flux divergence as the equivalent of absorption as far as the net vertical flux $\tilde{F}_z$ is concerned.
Equation (\ref{eqn:constitutive}) results directly from the diffusion approximation and acts as a constitutive law that closes the RT problem.
More specifically, it is the familiar Fick's law for the \emph{net} vertical vector flux, $\tilde{F}_z(z) = [-1/3(1-\omega\,g)\sigma]\,\tilde{J}^\prime(z)$, plus a source term.

BCs must also be Fourier transformed.
For that, we need the hemispherical fluxes along the $z$-axis $F_z^{(\pm)}$ from (\ref{eqn:F_z_up_down}) and, using (\ref{eqn:I_diff}) along with (\ref{eqn:J_def})--(\ref{eqn:Fz_def}), we obtain
\begin{equation}
\tilde{F}_z^{(\pm)}(z) = \frac{\tilde{J}(z)/2 \pm \tilde{F}_z(z)}{2}
\label{eqn:Ftilde_z_up_down}
\end{equation}
for the hemispherical fluxes.
In the diffusion approximation, we need to translate the radiance-based BCs in (\ref{eqn:pt_sc_and_refl})--(\ref{eqn:pt_sc_and_refl_opposite}) and (\ref{eqn:pt_sc_opposite_refl}) into flux-based statements using $\tilde{F}_z^{(\pm)}(z)$ at $z$ = 0 or $H$.

\subsection{Diffuse point-source on a reflective boundary}
\label{sec:diffuse_src}

This is the model used in \S\ref{sec:WISPR_footprint_theory} for the lower Rayleigh-scattering layer, leading to the expression for $\langle r^2 \rangle$ in (\ref{eqn:r2_PSP}).
The required BCs for the boundary-value problem based on the coupled ODEs in (\ref{eqn:conservation})--(\ref{eqn:constitutive}) are expressed in (\ref{eqn:pt_sc_and_refl})--(\ref{eqn:pt_sc_and_refl_opposite}) in physical $(x,y,z)$ space.
They Fourier transform to
\begin{align}
\label{eqn:pt_sc_and_refl_FT_I}
\tilde{I}_\text{bc}(\bOmega) &= \frac{1}{\pi}\,
                            \left[ 1 + \alpha\,\tilde{F}_z^{(-)}(0) \right],
                            \text{ at }z = 0,\text{ for }\Omega_z > 0 \\
\label{eqn:pt_sc_and_refl_opposite_FT_I}
\tilde{I}_\text{bc}(\bOmega) &= 0,\text{ at }z = H,\text{ for }\Omega_z < 0.
\end{align}

Multiplying  the above by $|\Omega_z|$ and integrating over the proper hemisphere leads to
\begin{align}
\label{eqn:pt_sc_and_refl_FT}
\tilde{F}_z^{(+)}(0) &= 1 + \alpha\,\tilde{F}_z^{(-)}(0), \\
\label{eqn:pt_sc_and_refl_opposite_FT}
\tilde{F}_z^{(-)}(H) &= 0.
\end{align}
Using (\ref{eqn:Ftilde_z_up_down}), we then get
\begin{align}
\label{eqn:pt_sc_and_refl_FT_diff}
\tilde{J}(0)+3\chi\tilde{F}_z(0) &= 4 + \alpha\,[\tilde{J}(0)-3\chi\tilde{F}_z(0)], \\
\label{eqn:pt_sc_and_refl_opposite_FT_diff}
\tilde{J}(H)-3\chi\tilde{F}_z(H) &= 0.
\end{align}

In these diffusion-friendly BCs, we have introduced, as is traditional in diffusion theory for the total (diffuse+direct) radiation, the so-called ``extrapolation'' scale factor $\chi$.
It is defined as the distance from the lower ($z = H$) boundary \emph{outside} the medium, measured in ``transport'' mean-free-paths (MFPs), $\ell_\rt = 1/(1-\omega\,g)\sigma$, where $\tilde{J}(z)$ vanishes in view of the lower BC (\ref{eqn:pt_sc_and_refl_opposite_FT_diff}) and Fick's law, $\tilde{F}_z(H) = -(\ell_\rt/3)\,\tilde{J}^\prime(H)$, relating net vertical flux to the local gradient in $\tilde{J}$.
The conventional value for $\chi$ is 2/3, but others have been used to improve the accuracy of the diffusion approximation compared to RT-based benchmarks \citep{bell1970nuclear}.
This helps to interpret physically the present type-3 (Robin) BCs as opposed to the more common type-1 (Dirichlet, specify $\tilde{J}$) or type-2 (Neumann, specify $\tilde{J}^\prime$) counterparts for the boundary-value problem at hand \citep{zwillinger2021handbook}.

Having a closed-form solution in hand for $\{\tilde{J},\tilde{F}_z\}(z)$, we are generally interested in the \emph{outgoing} hemispherical fluxes, viewed as a response to the imposed light source.
In the present case, we want (diffuse) reflectance $R$ and (total) transmittance $T$ in
\begin{align}
\label{eqn:dif_pt_sc_and_refl_Refl}
R &= [\tilde{J}(0)-3\chi\tilde{F}_z(0)]/4, \\
\label{eqn:dif_pt_sc_and_refl_Trans}
T &= [\tilde{J}(H)+3\chi\tilde{F}_z(H)]/4.
\end{align}
Adding these outcomes to the BCs in (\ref{eqn:pt_sc_and_refl_FT_diff})--(\ref{eqn:pt_sc_and_refl_opposite_FT_diff}), we get 
\begin{align}
1+R &= \tilde{J}(0)/2, \\
0+T &= \tilde{J}(H)/2,
\end{align}
where the additive terms on the left-hand sides are the normalized \emph{incoming} fluxes, thus relating $\{R,T\}(k)$ only to the values of the scalar flux $\tilde{J}(z)$ at the boundaries.

The present boundary value problem is easily set up in a symbolic math app such as Mathematica.
The resulting expressions for $R$ and $T$ are of most interest to us, as a function of all six parameters: $H$, $\omega$, $g$, $\sigma$ ($= \tau/H$), $\alpha$, and $k$ (along with $\chi$ = 2/3).
For instance, in (\ref{eqn:albedo}), we computed cloud albedo $\alpha_\rc = R = 1-T$ using a model configuration where there is forward-peaked ($g_\rc = 0.7$) conservative ($\omega_\rc = 1$) scattering with no boundary reflections ($\alpha = 0$) and uniform diffuse illumination ($k = 0$).
This yields $R = \tau_\rt/(2\chi+\tau_\rt)$, where the scaled (or ``transport'') cloud optical thickness $\tau_\rt = (1-g_\rc)\tau_\rc$ with $\tau_\rc = 25$ (cf. lower panel in Fig.~\ref{fig:AppB_Figs}).

Following \cite{davis2002space}, we then turned our attention in \S\ref{sec:diffusion_theory} to the lateral dispersion of the incoming adjoint/time-reversed NIR light in transmission. 
So, the configuration of the diffusion model implicit in (\ref{eqn:conservation})--(\ref{eqn:constitutive}), \emph{without} the source terms, subject to the BCs in (\ref{eqn:pt_sc_and_refl_FT_diff})--(\ref{eqn:pt_sc_and_refl_opposite_FT_diff}), has all the same parameter values, but now with $k > 0$.
By definition, $T(\tau_\rt/2\chi;k)/T(\tau_\rt/2\chi;0)$ is the AMTF for the cloud normalized, as is customary, to unity at $k = 0$.
\cite{davis2002space} then show that, by expanding the 
\begin{equation}
\text{AMFT}(k) = T(\tau_\rt/2\chi;k) \; \left/ \; T(\tau_\rt/2\chi;0) \right.
\label{eqn:AMFT_def}
\end{equation}
into a short Taylor series in $k$, up to order 2, one can infer the variance of $\rho$.
Recall that the inverse Fourier transform of AMTF$(k)$ is the APSF$(\rho)$, and it can be viewed as the probability density function (PDF) of the random horizontal distance $\rho$ to the point-source.
Therefore, we define
\begin{equation}
\langle \rho^2 \rangle = 2\pi \int\limits_0^\infty \rho^2 \text{APSF}(\rho)\rho\dif\rho \left/ 2\pi \int\limits_0^\infty \text{APSF}(\rho)\rho\dif\rho \right. ,
\label{eqn:VARrho_APSF}
\end{equation}
in case APSF$(\rho)$ is also normalized, as usual, to APSF$(0)$ = 1.
That way, solving APSF$(\rho)$ = 1/2 for $\rho$ yields the key FWHM property (divided by 2).

From that perspective, the AMTF (normalized to 1 at $k$ = 0), being the Fourier transform of the PDF of $\rho$, is known as its characteristic function in probability theory \cite{gnedenko1998theory}.
That means the statistical moments of $\rho$ can be obtained from successive derivatives of the AMTF at $k$ = 0.
In summary, we have
\begin{equation}
\text{AMTF}(k) = 1 + \frac{\langle \rho^2 \rangle}{2} \, k^2 + \mO(k^4).
\label{eqn:AMTF_series}
\end{equation}
In short, integration in physical space (\ref{eqn:VARrho_APSF}) is done by differentiation in Fourier space (\ref{eqn:AMTF_series}).
From there, we obtain
\begin{equation}
\langle \rho^2 \rangle/H_\rc^2 = \frac{2}{3} \, \left( 1 + \frac{3+4X}{2X(1+X)} \right),
\label{eqn:AMTF_VARrho}
\end{equation}
as stated in the main text, where $X = \tau_\rt/2\chi$ controls the pre-asymptotic correction term that decays as $1/X^{2}$ when $X \to \infty$.

A different version of this model is used in \S\ref{sec:WISPR_footprint_theory} to predict the contribution $\langle r^2 \rangle$ of the sub-cloud (Rayleigh scattering) layer to the total dispersion $(\langle \rho^2 \rangle + \langle r^2 \rangle)$ of the composite APSF for Venus' cloud-over-Rayleigh atmosphere.
In that case, the scattering is still conservative ($\omega_\rR = 1$) but now quasi-isotropic ($g_\rR = 0$, hence $\tau_\rt = \tau_\rR$) and the the boundary the light comes from is highly reflective ($\alpha_\rc = 0.85$).
In that case, the key property used in (\ref{eqn:AMFT_def}) will depend on $k$ and three parameters: $T(\alpha_\rc,\tau_\rt,\chi;k)$.
The evaluation in (\ref{eqn:AMTF_series}) is the same, but now yields $\langle r^2 \rangle$.
This leads to Eq.~(\ref{eqn:r2_PSP}) in the main text; the right-hand side of which comes numerically to 2.1 for these optical parameters (and $\chi$ = 2/3).
We note that its right-hand side can be expressed as a function of $X = \tau_\rt/2\chi$, as in (\ref{eqn:AMTF_VARrho}), and $\alpha_\rc$.

\subsection{Collimated point-source opposite a reflective boundary}
\label{sec:collimated_src}

This is the problem defined by the RT equation in (\ref{eqn:RTE}) for the diffuse radiance field, with the source term in (\ref{eqn:Q_for_collimated_src}), and BCs that are homogeneous on top (no incoming diffuse light) and reflective at the bottom, as captured in (\ref{eqn:pt_sc_opposite_refl}).
We will use it to model the role of the upper cloudy layer in the combined cloud/Rayleigh atmosphere.

\cite{Davis_etal2009} solved this problem in the diffusion limit for an absorbing ($\alpha = 0$) lower boundary; see their Appendix~C.
They also showed how to combine that solution with the one for an isotropic point-source that we used in the previous sub-section to account for the reflective (Lambertian) lower boundary; see their Eq.~(5.21) for $R(k)$.
A similar expression arises for $T(k)$, of immediate interest here, in the presence of multiple surface reflections.
However, \cite{Davis_etal2009} did not work out explicitly the combination of the medium's reflective and transmissive properties and the surface albedo.
Nor did they compute explicitly the lateral dispersion $\langle \rho^2 \rangle$ either for reflection or transmission, let alone in combination with a finite surface albedo ($0 < \alpha \le 1$), using the always-valid Taylor series approach in (\ref{eqn:AMTF_series}).

Here, we again set up the conservative ($\omega$ = 1) scattering model in Mathematica with the coupled ODE problem in (\ref{eqn:conservation})--(\ref{eqn:constitutive}), but this time \emph{with} the source terms in (\ref{eqn:Q_J})--(\ref{eqn:Q_Fz}), subject to homogeneous BCs (no need for the special $\chi$ parameter): $\tilde{J}(0)+2\tilde{F}_z(0) = \tilde{J}(H)-2\tilde{F}_z(H) = 0$.
The outgoing hemispherical fluxes of diffuse radiance are then: $R(k) = [\tilde{J}(0)-2\tilde{F}_z(0)]/4$ in reflection; and $T(k) = [\tilde{J}(H)+2\tilde{F}_z(H)]/4$ in transmission.
We can again simplify by bringing back the homogeneous BCs, ending up with: $R(k) = \tilde{J}(0)/2$ and, of prime interest here, $T(k) = \tilde{J}(H)/2$.

Using the same AMTF$(k)$ series expansion as before, this yields
\begin{equation}
\langle \rho^2 \rangle / H_\rc^2 = \frac{2}{3} \, \left( 1-\me^{-\tau} \right) \, \frac{ N_0(\alpha,g,\tau) + \me^{-\tau} N_1(\alpha,g,\tau) }{ D(\alpha,g,\tau) },
\label{eqn:VARrho_complex}
\end{equation}
where \\
$N_0(\alpha,g,\tau) = 15 (1-\alpha) (1-g)^2 \tau^3 + 60 (1-g) \tau^2$ \\ 
\indent\indent\indent\indent $+ 2g(65\alpha + 9(1-\alpha)(8-3g) - 25) \tau - 24 (5-3g)$, \\
$N_1(\alpha,g,\tau) = -6 (5\alpha-1) (1-g)^2 \tau^3 + 6(1-g)(-19\alpha-9g(1-\alpha) + 11) \tau^2$ \\ 
\indent\indent\indent\indent $-2(65\alpha - 9g(8\alpha + 3g(1-\alpha) - 12) - 85) \tau + 24 (5-3g)$,\\
and \\
$D(\alpha,g,\tau) = (1-g) \tau^2 ( 3 (1-\alpha) (1-g) \tau + 4 ) 
               \left( 5 - e^{-\tau} ( 3 \alpha (1-g) \tau + 1 ) \right).$ \\
Notice that the multipliers of the dominant ($\sim$$\tau^3$) terms in $N_0(\alpha,g,\tau)$ and in $D(\alpha,g,\tau)$ are identical: $15 (1-\alpha) (1-g)^2$.
Therefore, the asymptotic ($\tau \to \infty$) limit of (\ref{eqn:VARrho_complex}) is, as in (\ref{eqn:AMTF_VARrho}), 2/3, irrespective of $g$ and $\alpha$.

The cloud parameters from Fig.~\ref{fig:AppB_Figs} (lower panel) and \S\ref{sec:WISPR_footprint_theory} in the main text are, respectfully: $g_\rc$ = 0.7; $\tau_\rc$ = 25; and $\alpha_\rR$ = 0.73.
For the right-hand side of (\ref{eqn:VARrho_complex}), these numbers lead to 1.3, as stated in the main text.

\section*{Acknowledgments}

This research was carried out at the Jet Propulsion Laboratory, California Institute of Technology, under a contract with the National Aeronautics and Space Administration (80NM0018D0004). 
I thank Kevin Baines, Len Dorsky, James Cutts, Paul Byrne, and Nils Mueller (DLR) for fruitful discussions.

\bibliographystyle{elsarticle-harv} 

\bibliography{venus_imaging}







\end{document}